\documentclass{elsarticle}
\usepackage[utf8]{inputenc}
\usepackage{mathtools}
\usepackage{booktabs}
\usepackage{graphicx}
\usepackage{hyperref}
\usepackage{natbib}
\usepackage{siunitx}
\usepackage[a4paper,top=3cm,bottom=2cm,left=2.6cm,right=2.6cm,marginparwidth=1.75cm]{geometry}
\usepackage[final]{changes}

\journal{Renewable Energy}

\makeatletter
\def\ps@pprintTitle{
\def\@oddhead{
    \parbox[b][\headheight]{\textwidth}{
         \copyright ~2022. This manuscript version is made available under the CC-BY-NC-ND 4.0 license: \newline \url{https://creativecommons.org/licenses/by-nc-nd/4.0/}
    }}
\let\@evenhead\@empty
\def\@oddfoot{\footnotesize\itshape
    Article published in Renewable Energy. doi: \href{https://doi.org/10.1016/j.renene.2021.12.130}{10.1016/j.renene.2021.12.130}
    \hfill\today}
\let\@evenfoot\@oddfoot}
\makeatother

\begin{document}
\begin{frontmatter}
\title{Seasonal prediction of renewable energy generation in Europe based on four teleconnection indices}

\author[bsc]{Llorenç Lledó\corref{cor1}}
\ead{llledo@bsc.es}
\author[bsc]{Jaume Ramon}
\author[bsc]{Albert Soret}
\author[bsc,icrea]{Francisco-Javier Doblas-Reyes}
\address[bsc]{Barcelona Supercomputing Center (BSC)}
\address[icrea]{ICREA, Pg. Lluís Companys 23, Barcelona 08010, Spain}
\cortext[cor1]{Corresponding author}

\begin{abstract}

With growing amounts of wind and solar power in the electricity mix of many European countries, understanding and predicting variations of renewable energy generation at multiple timescales is crucial to ensure reliable electricity systems. At seasonal scale, the balance between supply and demand is mostly determined by the large-scale atmospheric circulation, which is uncertain due to climate change and natural variability.
Here we employ four teleconnection indices, which represent a linkage between atmospheric conditions at widely separated regions, to describe the large-scale circulation at seasonal scale over Europe.
For the first time, we relate each of the teleconnections to the wind and solar generation anomalies at country and regional level and we show that dynamical forecasts of the teleconnection indices allow predicting renewable generation at country level with positive skill levels.
This model unveils the co-variability of wind and solar generation in European countries through its common dependence on the general circulation and the state of the teleconnections.

\end{abstract}

\end{frontmatter}


\section{Introduction}
The European Commission is pursuing that Europe becomes the first climate-neutral continent by 2050, and it has proposed to cut down green-house gas (GHG) emissions to at least 50\% of 1990 levels by 2030 \citep{EGD2019}. In order to do so, a decarbonisation of the energy sector ---which is still the largest contributor to GHG emissions in the continent--- is taking place with massive installation of wind and solar power capacity. 
But wind and solar power plants are weather-dependent and non-dispatchable sources (i.e. its production is intermittent and non-programmable). 
This draws a future European context with a high penetration of variable renewable energy (VRE) generation in the electricity mix and a growing role of the atmospheric circulation in shaping electricity generation. Therefore, the need for accurate prediction of renewable energy generation at multiple timescales is also a growing concern for many actors in the electricity system \added{\citep{irena2020}}. 
In particular, forecasts of VRE generation at country level can serve transmission system operators to schedule alternative power sources, \deleted{and} energy traders to anticipate electricity prices\added{, and governments to prevent electricity price crises}.\\ 

\replaced{Weather forecasts have been long used in the sector to anticipate energy generation variability from hours up to ten days ahead }{At short timescales (from hours up to ten days ahead), weather forecasts are routinely used and cope with this challenge} with remarkable success \citep{Alessandrini2017, Giebel2017}. But beyond that scale, the chaotic nature of the atmosphere \replaced{turns }{makes} weather forecasts useless, and climatological estimates are typically used to plan activities or to estimate future risks. However, new developments in the field of climate prediction are making \added{probabilistic} forecasts at longer timescales (i.e. sub-seasonal, seasonal or decadal predictions) possible \citep{Meehl2021, Merryfield2020}.
Indeed, \deleted{seasonal} forecasts of \added{cumulative} energy generation and demand \added{during a whole season} are not only possible but highly needed for the energy sector \citep{Orlov2020, Soret2019, Lledo2019}. \added{While seasonal forecasts of wind power generation at wind farm level have already been studied in \citet{Lledo2019} and those of solar generation were studied in \citet{DeFelice2019}, this work focuses on producing and analyzing country-wise VRE generation forecasts at seasonal timescales in Europe, which poses specific challenges that have not been addressed in the literature as far as the authors know.}\\ 

\added{
The year-to-year variations of seasonal VRE generation are mostly driven by the state of the climate system in each particular year.
Long-lived anomalies of slowly-varying fields such as tropical sea surface temperatures \citep{Lledo2018}, Arctic sea ice extent \citep{Acosta2019}, snow cover, or soil moisture \citep{Prodhomme2015} force the atmosphere towards anomalous conditions \citep{Shukla2006} and end up impacting VRE generation. 
While it is impossible to determine precisely the time evolution of atmospheric variables during a season, preferred states (e.g. likelihood of experiencing above normal or below normal conditions) can be estimated with an ensemble of Earth System Model simulations initialized at the beginning of the season with slightly different but equally plausible initial conditions. Each initial condition results in a different trajectory (known as ensemble member), and a statistical analysis of their distribution reveals the preferred states in form of probabilistic forecasts \citep{DoblasReyes2013}.}\\

However, seasonal forecasts of atmospheric variables are produced at very coarse resolutions, and are affected by large biases. Therefore, statistical post-processing methods such as calibration \citep{Torralba2017}, downscaling \citep{Ramon2021} or bridging \citep{Specq2020} are needed at some point to translate atmospheric variables to energy quantities. Particularly, bridging methods allow \deleted{to} transform\added{ing} forecasts of \deleted{a} teleconnection \replaced{indices}{ index} ---\replaced{quantities that summarize the state of the climate during a particular season over well-defined large areas}{that describes preferred states of the climate system impacting on well-defined large areas}--- \replaced{into }{onto} another variable of interest \added{(country-wise cumulative VRE generation during a season in our case)}, if a robust link between them is found in observational records (e.g. \citet{Lledo2020b}). 
The rationale behind bridging methods is that when the variable of interest can be linked to large-scale circulation indices, there is no need to represent the local scales in the dynamical models, and current coarse-scale predictions are sufficient if the large-scale circulation is appropriately represented.\\

The general atmospheric circulation over Europe at seasonal timescales can be described by four teleconnection indices: the North Atlantic Oscillation (NAO) \citep{Hurrell2003}, the East Atlantic (EA) \citep{Woollings2010}, the East Atlantic/Western Russia (EAWR) \citep{Lim2014} and the Scandinavian pattern (SCA) \citep{Bueh2007}.
The state of those Euro-Atlantic teleconnections determines the strength and location of the jet stream (e.g. \citet{Woollings2010}) and affects near-surface wind speed and solar radiation in Europe \citep{Jerez2013a, Brayshaw2011, Zubiate2017, Correia2020, Jerez2013b}.
By using pseudo-observations of VRE generation at country and regional (NUTS2) level in Europe \added{from state-of-the-art energy reanalyses}, we determine the impact of each teleconnection in VRE generation and show that, in some cases, those four indices describe most of the generation  variability at country level. Then, using a multi-linear regression (following \citet{Ramon2021}), we produce and evaluate the quality of seasonal forecasts of \added{country-wise} wind and solar power generation by bridging seasonal climate predictions of the teleconnections derived from five Seasonal Prediction Systems (SPS), that were shown to have some skill levels in \citet{Lledo2020}.

\section{Methods}
\label{sec:methods}

\subsection{Pseudo-observations of wind and solar power generation}
\label{sec:cfobs}
Many European countries provide observational records of national power generation by source type. However, the datasets are only available from sparse sources and the data is not homogeneous in terms of temporal resolution, the variables provided, and even the definitions employed to group by generation source \citep{Wiese2019}. Although ENTSO-e provides unified access to data from 35 European countries \citep{Hirth2018}, its transparency platform only contains the records from 2015 onward, which is not sufficient for long-term analyses. Additionally, to produce comparable results, the generation data needs to be scaled with installed capacity, which is rarely available at the same time resolution as the generation data itself. Even with huge data-rescue and quality control initiatives such as the open-power-system-data \citep{Wiese2019}, the observational datasets are not fit for the purpose of this work.\\

Alternatively, estimated values of aggregated wind and solar capacity factors (CF) at country level have been obtained from three pseudo-observational datasets: the EMHIRES dataset \citep{EmhiresWind,EmhiresSolar}, the NINJA dataset \citep{Staffell2016,Pfenninger2016} and the UREAD-ERA5 Energy Reanalysis dataset \citep{UREADERA5}. 
Those three datasets estimate CF values using a bottom-up physical approach: sub-daily wind speed, radiation and temperature values are obtained at the VRE fleet locations from a reanalysis dataset or a satellite product. Then the (sometimes limited) available information of each power plant performance is used to estimate CF values, which are then averaged at regional/country level. 
This conversion from atmospheric to energy variables uses a fixed power plant fleet, thus resulting in hypothetical CF values that would have been obtained in the past years with the specified fleet. 
Specifically, EMHIRES assumes a fixed fleet that represents the existing installations in 2014, UREAD-ERA5 uses the 2017 fleet, and NINJA uses a fleet from December 2016 (see table~\ref{tab:CF_datasets}). 
Although this physical approach has many limitations (lack of spatial resolution, lack of plant performance information, etc...), it has the advantage of producing long and temporally-consistent CF datasets that reflect only the variability of the atmospheric conditions. Those are desirable properties for training and evaluating statistical models and for extracting robust conclusions. \\

\begin{table}[ht]
\centering
\caption{Summary of the CF datasets employed.}
\label{tab:CF_datasets}
\begin{tabular}{lcccccc}
\toprule
 & EMHIRES & NINJA & UREAD-ERA5 \\
\midrule
Period & 1986--2015 & \begin{tabular}{@{}c@{}} 1980--2016 (wind) \\ 1986--2016 (solar) \end{tabular} & 1979--2018 \\
Fleet & 2014 & 2016 & 2017 \\
Wind model inputs & MERRA2 & MERRA2 & ERA5 \\
Solar model inputs & SARAH & SARAH + MERRA2 & ERA5 \\
\bottomrule
\end{tabular}
\end{table}

The works of \citet{Moraes2018} and \citet{Kies2021} have intercompared several aspects of these datasets and found significant differences in the provided CF values. However, it is difficult to determine which product has the best overall quality, because the usage of different fixed fleets makes comparisons to actual CF data unfair. For each dataset, only the year that corresponds to the fleet employed can be faithfully verified, and that year changes from one dataset to another. Moreover, those verifications are typically made at the hourly or daily scale, whereas for checking interannual variability ---which is the key element in this work--- , consistent records over many years are not available.
In this work the EMHIRES dataset has been used as the main reference dataset, while keeping the other two sets for evaluating the sensitivity of the results to this choice. \\

For the three datasets, seasonal mean values of CF at country level have been computed from hourly values for the four seasons of the year (DJF, MAM, JJA, SON). Additionally, the EMHIRES dataset provides regional data aggregated at NUTS2 level (the second administrative level of the European nomenclature of territorial units for statistics), which is used to understand the role of spatial aggregation at country level. 

\subsection{Observations of Euro-Atlantic teleconnection indices}
\label{sec:eatcobs}
Observed teleconnection indices of the NAO, EA, EAWR and SCA for the four seasons of the year have been derived from the ERA5 global reanalysis \citep{ERA5, Hersbach2020} for the period 1979--2018. The indices are computed by performing a Rotated Empirical Orthogonal Function (REOF) analysis of \SI{500}{hPa} geopotential height (Z500) seasonal mean anomalies, as described in \citet{Lledo2020}. 
This method can be seen as a dimensionality reduction technique that approximates the seasonal mean anomalies of Z500 for a given year as a linear combination of four fixed patterns, reducing drastically the collinearity in the grid-point data and filtering out small-scale variability.
However, note that due to the Varimax rotation employed, the teleconnection indices still have some degree of correlation between them.\\

\subsection{Seasonal hindcasts of Euro-Atlantic teleconnection indices}
\label{sec:eatchcsts}

Ensemble seasonal predictions of the four Euro-Atlantic teleconnection indices have been obtained from five different Seasonal Prediction Systems (SPS) that are made available by the Copernicus Climate Change Service (C3S): the System2 from Deutscher Wetterdienst (DWD2, \citet{DWD}), the GloSea5-GC2 (GSCGC2) from the UK Met Office \citep{Maclachlan2015, Williams2015}, the System 6 from Météo France (MF6, \citet{Dorel2017}), the SEAS5 \citep{Johnson2019} from the European Centre for Medium-Range Weather Forecasts and the Seasonal Prediction System 3 (SPS3) from Centro Euro-Mediterraneo sui Cambiamenti Climatici \citep{Sanna2017}. 
The C3S provides hindcasts (i.e. retrospective predictions) produced with the same state-of-the-art prediction systems that are used for operational prediction. They cover the 1993--2016 period in a unified 1ºx1º grid for all the systems, and serve to assess the forecast quality \citep{Hamill2006, Takaya2019}.
Additionally, a multi-system combination was also produced by pooling all the ensemble members together. Such multi-system combinations benefit from the strengths of the different contributors and produce more robust results than the individual systems \citep{Hagedorn2005}, and should therefore be preferred. \\

For each prediction system, ensemble member and year, the teleconnection forecasts are obtained by projecting the Z500 seasonal mean anomalies onto the ERA5-derived teleconnection patterns. A more detailed description of this process can be found in \citet{Lledo2020}, which also presents a comprehensive evaluation of the quality of the teleconnection index forecasts.

\subsection{Deriving energy production from the teleconnection indices}
\label{sec:stat_method}

VRE generation in European countries can be modelled through the state of the four aforementioned teleconnection indices. For each country or NUTS2 region under consideration, we build a statistical model that relates observed values of seasonal-mean CF to the state of the four teleconnection indices during that season: 
\begin{equation}
\label{eq:1}
    CF \approx f(NAO, EA, EAWR, SCA)
\end{equation}
The most suitable function $f$ to approximate the actual CF values is to be found by analyzing the contemporaneous records of CF and teleconnection index observations. For the EMHIRES dataset this results in a period of 30 years, and given that the analyses are carried out at seasonal timescales and separately for each season, the sample sizes consist of only 30 elements for estimating the model parameters. Therefore, only simple statistical models with few parameters can be employed. Accordingly, a multi-linear regression with a pre-selection of the relevant predictors is fit to the historical data for each region: 
\begin{equation}
\label{eq:2}
    CF(t) \approx a \cdot NAO(t) + b \cdot EA(t) + c \cdot EAWR(t) + d \cdot SCA(t)
\end{equation}
The parameters are estimated by least squares, but the pre-selection of the predictors fixes some of the parameters $a$, $b$, $c$ or $d$ in eq.~\ref{eq:2} to zero and serves to prevent collinearity and overfitting. A forward and backward stepwise selection based on Akaike Information Criteria (AIC) is employed to such end \citep{James2013}. 
\added{Note that the selection of the predictors does not take into account the accuracy of the teleconnection forecasts employed later on.}\\

Given that in this setting the predictors are obtained from a REOF analysis of Z500, the overall method can be seen as a Principal Component Analysis regression \citep{Alonzo2020, James2013}.
The goodness of fit (i.e. accuracy of the approximation) has been measured by the determination coefficient $r^2$ between fitted values (right side of eq.~\ref{eq:2}) and actual values of CF (left side of eq.~\ref{eq:2}). 

\subsection{Transforming teleconnection forecasts into energy forecasts}

The statistical model developed in the previous paragraphs can be employed to transform (or bridge \citep{Specq2020}) seasonal predictions of teleconnection indices into country or NUTS2 CF forecasts ($\widehat{CF}$), just by plugging predicted values of the teleconnection indices ($\widehat{NAO}$, $\widehat{EA}$, ...) into the function $f$: 
\begin{equation}
\label{eq:3}
    \widehat{CF} \coloneqq f(\widehat{NAO}, \widehat{EA}, \widehat{EAWR}, \widehat{SCA})
\end{equation}

The proposed setting (known as perfect prognosis \citep{Marzban2006}) does not correct any potential biases in the teleconnection indices used as predictors, but has two main advantages: firstly, the usage of observations for training the statistical model allows for larger training samples (i.e. we are not restricted to the teleconnection hindcast length, which consists of 24 years only). Secondly, the statistical model can be reused for any prediction of the indices despite its origin (i.e. we can use teleconnection forecasts from any seasonal prediction system or even from empirical methods). \\

The procedure is done individually for each ensemble member, producing an ensemble of CF forecasts that can be used to estimate probabilities of certain events. Following standard practice in the climate prediction arena, the ensemble members are distributed in three categories (below normal CF/normal CF/above normal CF) that are defined based on the terciles of the historical distribution of CF values. Then the probability of observing each tercile category is derived from the proportion of ensemble members in each group \added{(see figure~\ref{fig:fcst_example} for an example)}.

\subsection{Forecast verification}
\label{sec:verif}
The quality of the CF forecasts has been evaluated for the hindcast covering the years 1993--2016.
The EMHIRES CF dataset that was employed to fit the statistical modelling has also been used as verification truth.
In order not \added{to} use the verifying observations during the statistical model training step, the verification has been done in a leave-one-out cross validation setting \citep{James2013}. For each year that we want to predict and verify, one instance (seasonal mean value of CF) is kept aside for verification and a separate model is trained with the remaining observations.\\

The probabilistic CF forecasts have been verified with three selected metrics that measure different quality aspects \citep{Jolliffe2011}. First, Ensemble Mean Correlation (EMC) is a very useful indicator of potentially available skill. If a forecast does not contain a minimum level of EMC it will be very difficult to extract any useful information from it, therefore it is a good choice for a first screening.
Secondly, the Ranked Probability Skill Score (RPSS) measures the accuracy of forecasts presented in the form of tercile probabilities (i.e. probabilities of observing below normal, normal or above normal CF), when compared to climatological probabilities for those three categories (33\% each).
RPSS takes into account both the sharpness and the reliability of predictions, and is known to be a harsh standard \citep{Mason2004}.
In third place, the discrimination for the upper and lower tercile forecasts is evaluated by means of Relative Operating Characteristic (ROC) curves, which display the hit rate and false alarm rate at various probability thresholds. For each ROC curve, the ROC Skill Score (ROCSS) is obtained by computing the area under the curve (AUC) and comparing it to the value of 0.5 that would be obtained with random forecasts (i.e. $ROCSS = 2 \cdot AUC - 1$).

\section{Results}
\label{sec:results}

\subsection{Impact of the teleconnections on wind and solar generation}
\label{sec:impact}

\begin{figure}[ht!]
\includegraphics[width=0.95\textwidth]{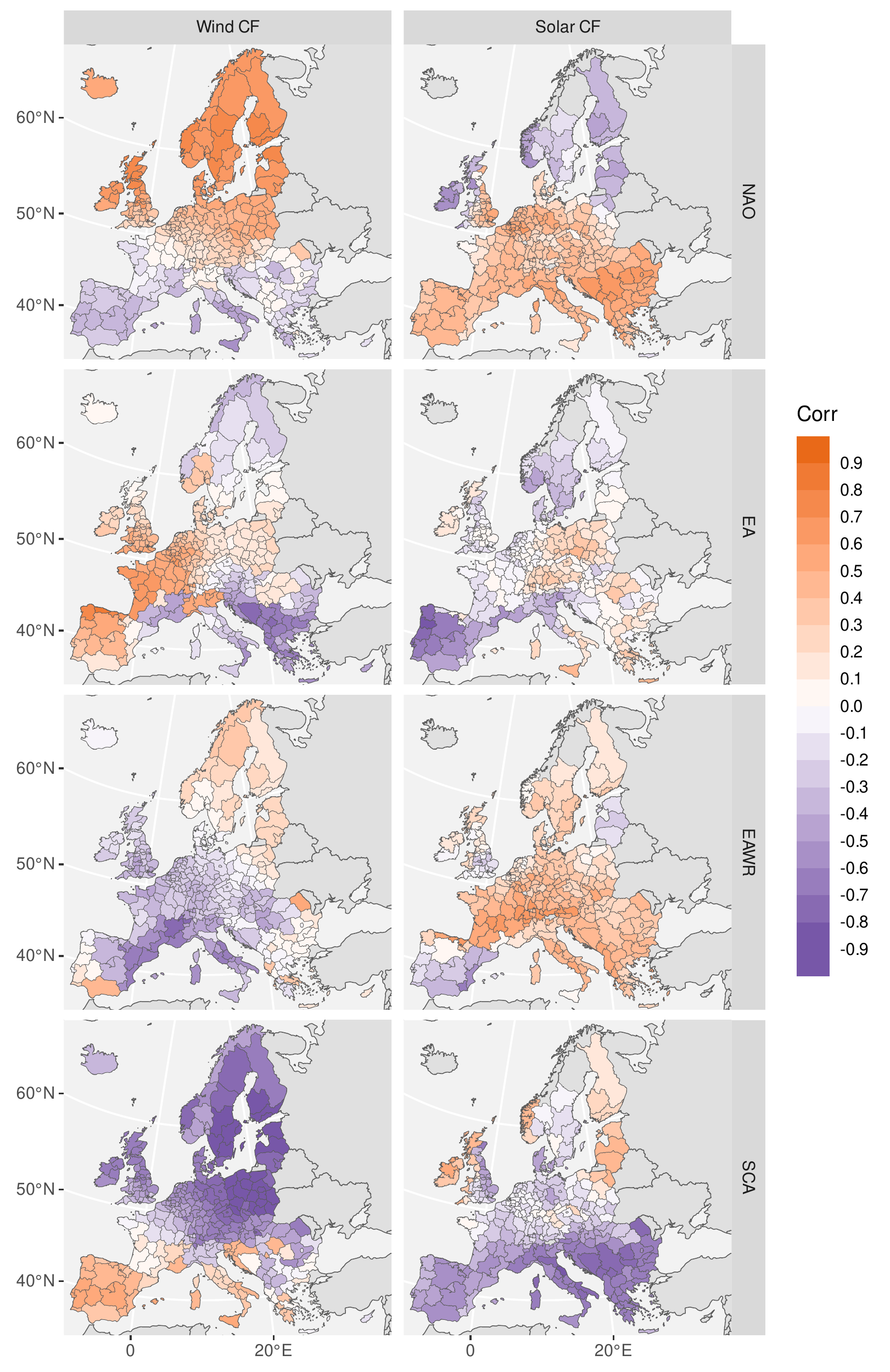}
\caption{Pearson correlation at each NUTS2 European region between the four observed (ERA5) teleconnection indices (rows) and the observed wind (left column) or solar (right column) CF values from the EMHIRES dataset, for DJF seasonal averages in the 1986/87--2014/15 period.}\label{fig:corr_cf_nuts2}
\end{figure}

The impact of each teleconnection on wind and solar generation has been evaluated by correlating the historical values of the teleconnection indices to the capacity factor (CF) pseudo-observations at country and regional level. 
This correlations illustrate that knowledge of each teleconnection state is useful to describe the anomalies of wind and solar generation in different European regions.
In winter (figure~\ref{fig:corr_cf_nuts2}), the positive phases of the NAO are linked to increased wind power generation north of 50ºN of latitude, and to reduced generation in Spain and south of Italy.
The EA has its strongest impacts on those regions that the NAO does not influence, increasing wind power generation in the central Atlantic façade during positive phases, and decreasing it in the Balkans.
The negative phases of the EAWR are mainly associated to north-easterly flows in the western Mediterranean region (i.e. Mistral channeling in the Ebro valley and the Gulf of Lion). The impact of the SCA resembles the reversed impact of the NAO, but a bit shifted towards the southeast. This is no surprise, since NAO and SCA have a correlation of -0.52 in winter.\\

For solar power, positive NAO phases increase generation in southern and central Europe, especially in the Balkans. The EA positive phases reduce generation in the Iberian peninsula, particularly in Portugal. Positive EAWR phases increase solar generation in central Europe, especially in France, Switzerland and Germany, but also in the northwest of Spain, parts of Italy and the Balkans. The impact of the SCA also looks like a southward shifted and reversed impact of the NAO, with its strongest signal over Italy and the Balkans. \\

\begin{figure}[ht!]
\includegraphics[width=\textwidth]{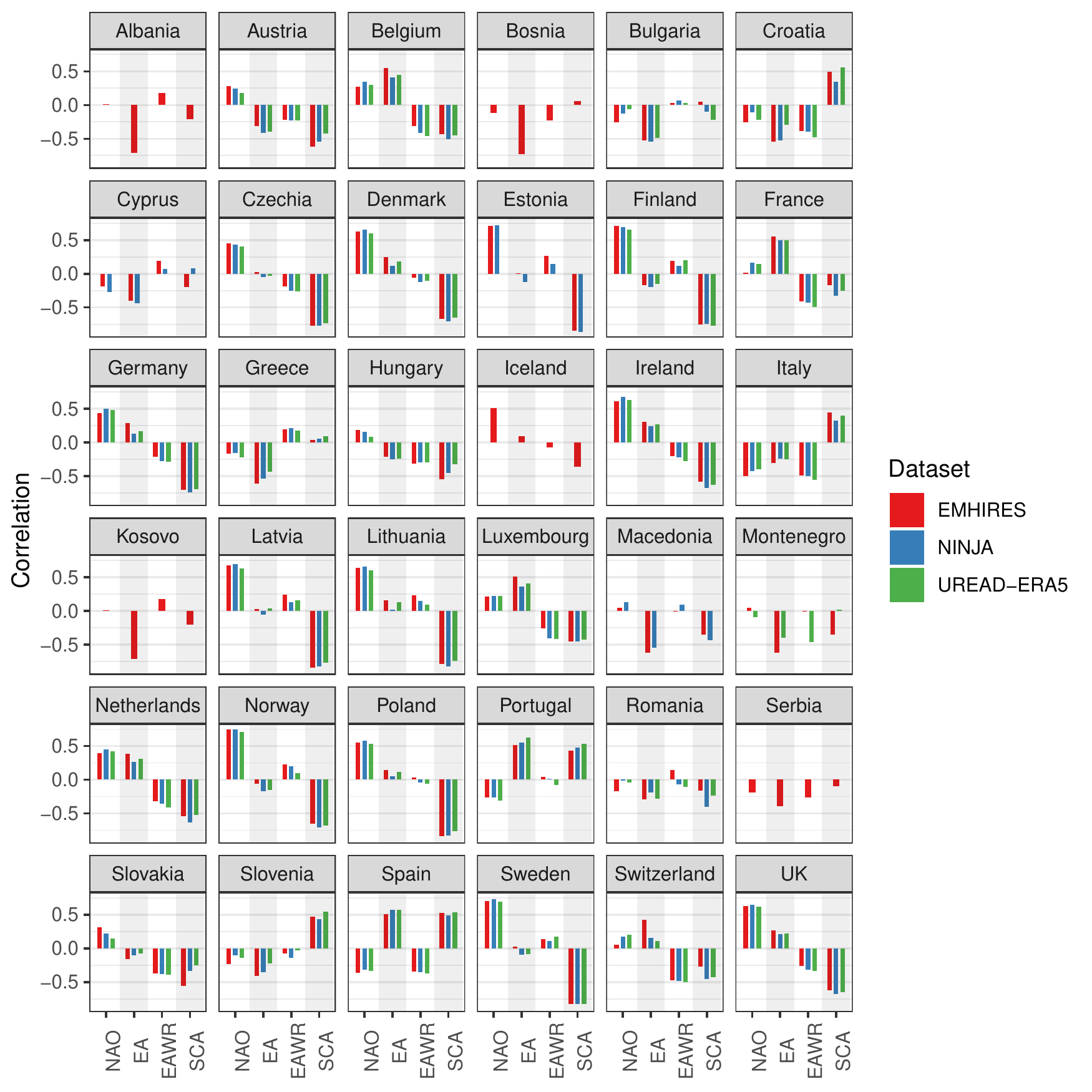}
\caption{Correlation coefficients between observed teleconnection indices and estimated wind CFs \added{in winter, obtained} from three datasets \added{and} for each of the European countries.}\label{fig:corr_cf_country}
\end{figure}

A similar analysis at country scale (figure~\ref{fig:corr_cf_country}) reveals that in many countries not only the NAO ---which has been the target of many studies--- is relevant to determine wind production. In Spain, the state of the EA and the SCA have more impact that the NAO. In France, the EA is the teleconnection with most impact, even if the regional impacts (figure~\ref{fig:corr_cf_nuts2}) are of opposite sign in the southeast of the country. And in Italy, the EAWR is the most relevant teleconnection.\\

For each European country, the correlations have been computed from the three CF datasets (see Methods) to evaluate the sensitivity to the observational reference. Despite the differences between the three datasets in terms of the available period, the power plant fleet and the energy conversion methodology employed, the correlations between the teleconnection indices and the country-aggregate CF values remain relatively stable. The largest deviations are seen in Montenegro, where the renewable fleet is small, and therefore minor differences in the considered wind farm locations can result in substantial CF differences. 

\clearpage

\subsection{Explanatory power of the teleconnections}

\label{sec:predictpower}
\begin{figure}[ht!]
\includegraphics[width=1\textwidth]{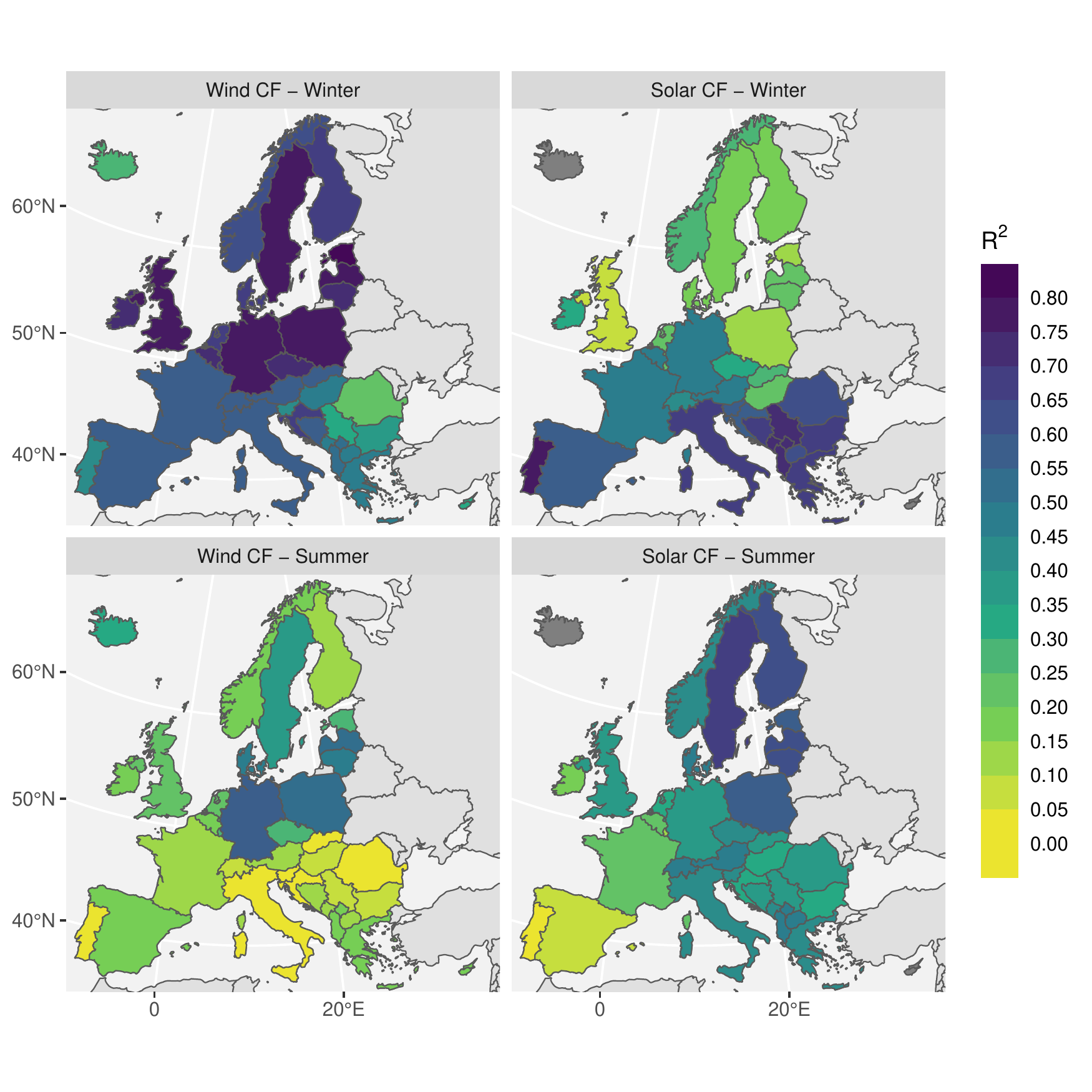}
\caption{Determination coefficient ($r^2$) of the multi-linear regression model for wind (left column) and solar (right column) CFs at country level in winter (top row) and summer (bottom row).}\label{fig:rsquared}
\end{figure}

A multi-linear regression between the observed teleconnection indices and the historical capacity factors has been fitted for each European country (see Methods). 
The determination coefficient $r^2$ of each regression indicates the goodness of fit of the statistical model, or equivalently, the percentage of CF variance that can be explained by the linear combination of the teleconnection indices. 
Figure~\ref{fig:rsquared} shows a map of the determination coefficients for wind and solar generation in winter and summer. 
For wind CFs, the $r^2$ is high in winter and rather low in summer in most European countries. The teleconnection indices explain most of the variability of winter wind power generation in the UK, Ireland, Germany, Poland, Sweden and the Baltic countries. Regarding solar CFs, the teleconnections have good predictive ability for southern Europe in winter, whereas in summer Sweden, Finland, the Baltic countries and Poland also have good $r^2$ values. \\

\added{As explained in section~\ref{sec:stat_method}, a pre-selection of predictors was used to prevent overfitting. Table~\ref{tab:reg_coeffs} presents the regression coefficients ($a$, $b$, $c$ and $d$ in equation~\ref{eq:2}) obtained for wind and solar CFs in winter and summer for four relevant countries. These regression coefficient indicate how much the CF changes when a teleconnection index experiences a one-standard-deviation change. We can see that in many cases the pre-selection discards some predictors. For instance, winter wind capacity factors in Spain are modelled as a function of EA, EAWR and SCA only, discarding NAO as a predictor. This does not imply that NAO has no effect on wind CFs in Spain, but rather than the information that the NAO index provides is redundant once we know the values of the other three indices.} 

\begin{table}[ht]
\centering
\caption{\added{Regression coefficients obtained when modelling wind and solar capacity factors as a function of the four teleconnection indices for Germany, UK, France and Spain in winter and summer. NA values indicate that the predictor was not included for that country. The determination coefficient $r^2$ is also indicated.}}
\label{tab:reg_coeffs}
\begin{tabular}{lcccccccccc}
\toprule
        & \multicolumn{5}{c}{Wind CF}                 & \multicolumn{5}{c}{Solar CF}                \\
        \cmidrule(lr){2-6} \cmidrule(l){7-11}
        & NAO     & EA     & EAWR    & SCA     & $r^2$   & NAO    & EA      & EAWR    & SCA     & $r^2$   \\
\midrule        
Winter  &         &        &         &         &      &        &         &         &         &      \\

\quad Germany & NA      & 0.025  & -0.026  & -0.055  & 0.76 & 0.0030 & NA      & 0.0025  & 0.0013  & 0.47 \\
\quad UK      & 0.022   & 0.021  & -0.024  & -0.032  & 0.77 & 0.0009 & NA      & NA      & NA      & 0.08 \\
\quad France  & NA      & 0.030  & -0.025  & -0.017  & 0.58 & 0.0032 & NA      & 0.0038  & NA      & 0.47 \\
\quad Spain   & NA      & 0.030  & -0.017  & 0.027   & 0.55 & 0.0025 & -0.0037 & NA      & -0.0039 & 0.59 \\
\\
Summer  &         &        &         &         &      &        &         &         &         &      \\
\quad Germany & -0.0039 & 0.0058 & -0.0110 & NA      & 0.58 & 0.0031 & 0.0023  & 0.0019  & 0.0018  & 0.35 \\
\quad UK      & NA      & 0.0078 & -0.0062 & NA      & 0.24 & 0.0047 & -0.0028 & NA      & NA      & 0.40 \\
\quad France  & -0.0066 & NA     & NA      & NA      & 0.11 & 0.0025 & NA      & NA      & 0.0021  & 0.21 \\
\quad Spain   & NA      & NA     & NA      & -0.0079 & 0.18 & NA     & NA      & -0.0012 & NA      & 0.07 \\
\bottomrule
\end{tabular}
\end{table}

\subsection{Potential value of wind and solar generation forecasts}
\label{sec:screening}
The work of \citet{DeFelice2019} introduces a framework to evaluate the usefulness of seasonal predictions of renewable generation for a region in terms of three elements: a) the VRE installed capacity in the region; b) the amounts of interannual variability in the resource; and c) the forecast quality. Accurate forecasts would be useless in regions where there is no installed capacity, or where the variability is small. The two first aspects can be evaluated from installed capacities and CF observations alone and are evaluated in this section, whereas the forecast quality is evaluated in the next section.\\

\begin{figure}[ht!]
\includegraphics[width=\textwidth]{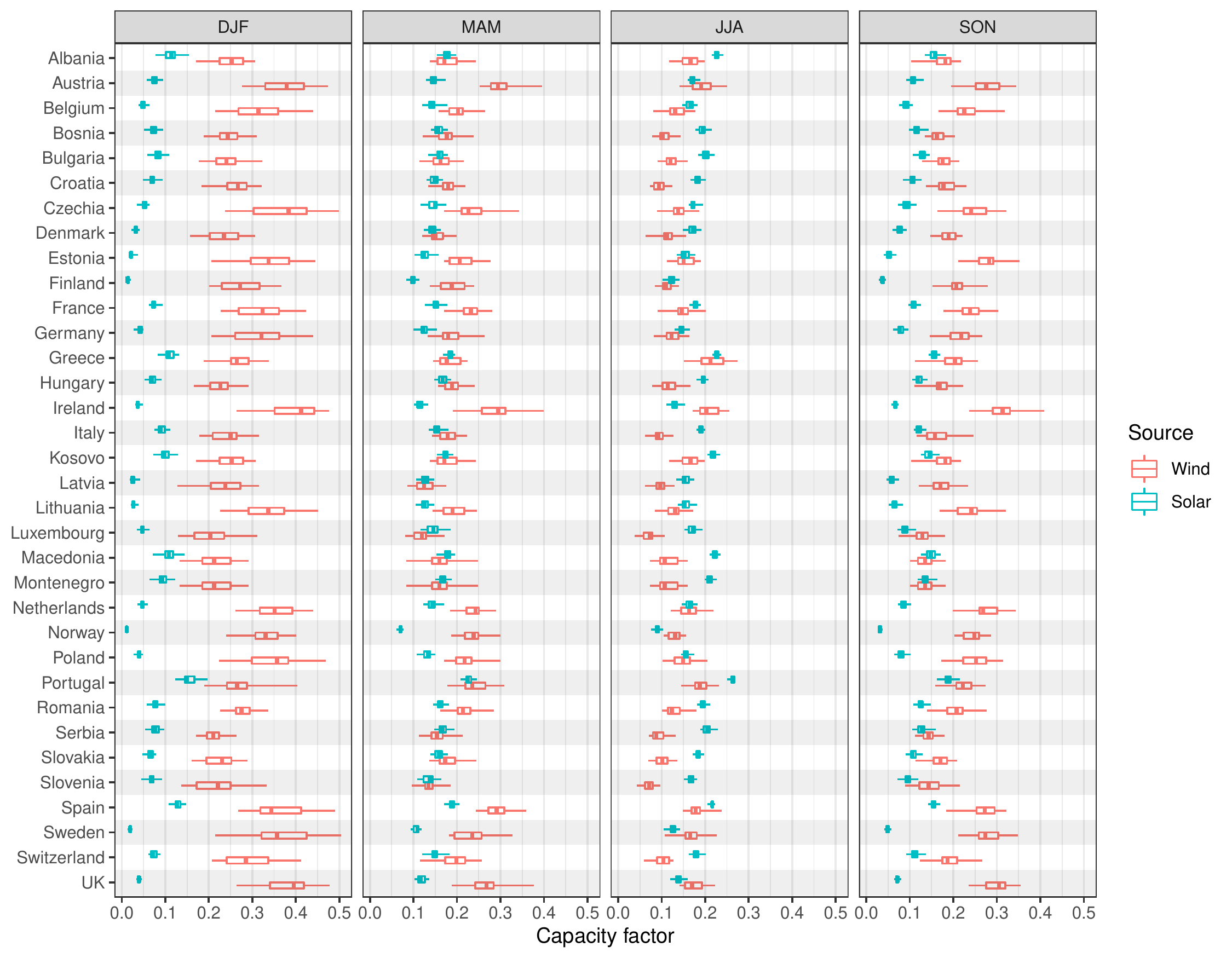}
\caption{Interannual variability of wind (red) and solar (blue) seasonal-mean capacity factors in the 1986--2015 period for 34 European countries, according to the EMHIRES dataset. Each boxplot represents the 30 seasonal-mean values for a given season, generation type and country. The whiskers depict the range (min-max) and the boxes the three quartiles.}\label{fig:variability_cf}
\end{figure}

For a), according to records of national installed capacity in 2019 \citep{OPSD_InstalledCapacity}, Germany, Spain, UK and France are the four countries with more VRE installed capacity. In these countries, as in most of the other European countries, installed wind power capacity more than doubles installed solar capacity.
Regarding b), the interannual variability of solar and wind resource in Europe is evaluated in figure~\ref{fig:variability_cf} for each season. The year-to-year variability of wind CFs is larger than solar variability for all the countries and seasons. Although solar CF has a strong seasonal cycle (driven by astronomic changes in the solar angle), its interannual variability does not change substantially from one season to another. Contrarily, wind power generation is much more variable in winter than in summer, with autumn and spring having intermediate variability levels.\\

Therefore, from a European perspective, forecasting winter means of wind power CF in the four aforementioned countries would be the most useful in terms of the total VRE generation that can be anticipated. 

\subsection{Bridging teleconnection forecasts to generation forecasts}

\label{sec:forecasts}
\begin{figure}[ht!]
\includegraphics[width=\textwidth]{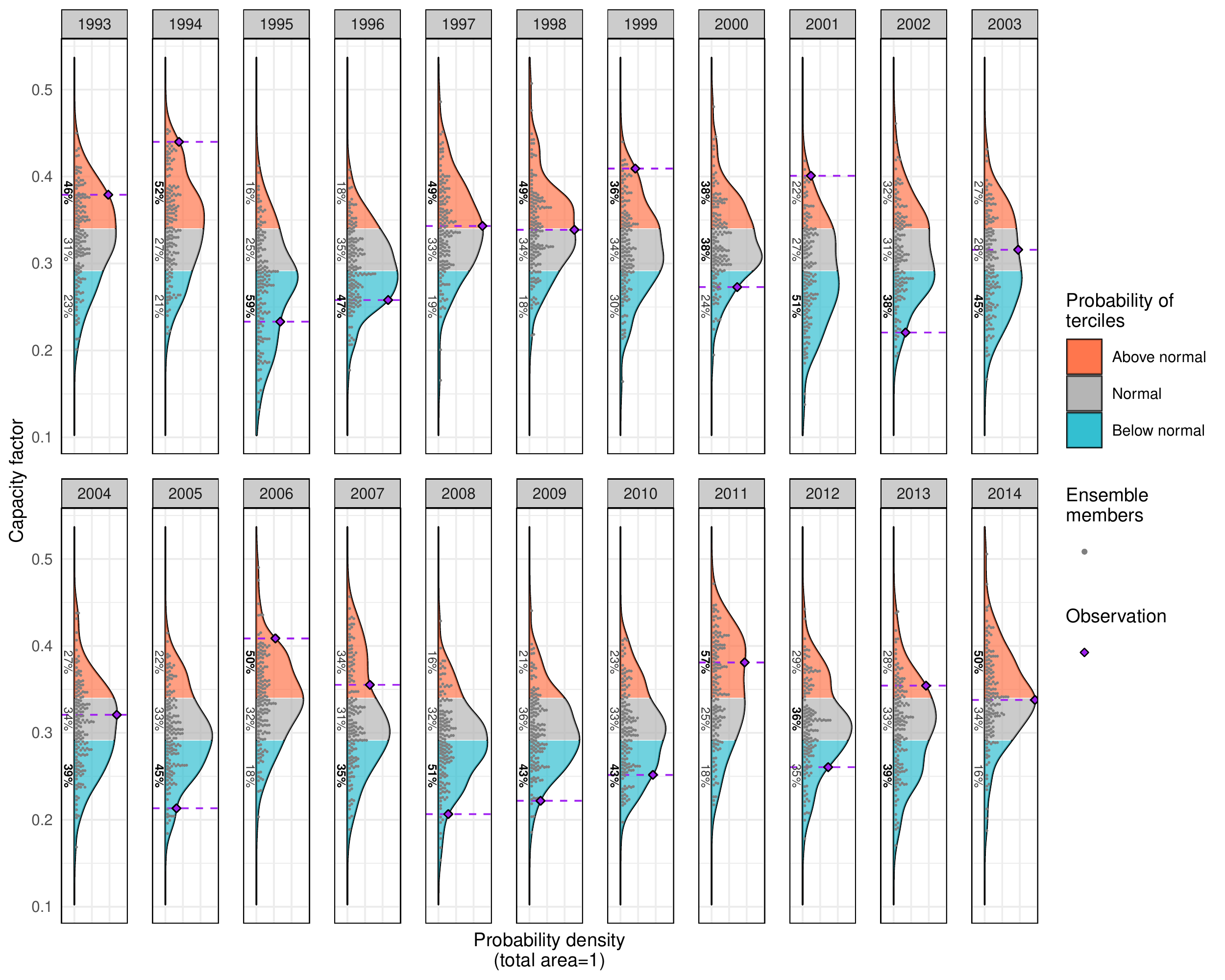}
\caption{\added{Retrospective forecasts of wind capacity factor in Germany, issued on December 1\textsuperscript{st} and valid for DJF. Each panel shows the multi-system predictions (148 members, grey dots) for a specific year, and the corresponding observed value. Probabilities of above normal/normal/below normal capacity factor conditions are labelled at the side of each panel. The most likely tercile is emphasized in bold. The year indicated refers to the beginning of the season.}}\label{fig:fcst_example}
\end{figure}

The multi-linear regression models that were fitted with observations are now directly fed with seasonal forecasts of teleconnection indices to obtain CF forecasts at country level. 
As an example, figure~\ref{fig:fcst_example} displays the probabilistic wind capacity factor forecasts of the multi-system combination for Germany in the winters of 1993/94 to 2014/15, initialized at the beginning of the season. For this country and season, the capacity factor forecasts were obtained as:
\begin{equation}
\label{eq:4}
\widehat{CF} = 0.32 + 0.025 \cdot \widehat{EA} - 0.026 \cdot \widehat{EAWR} - 0.055 \cdot \widehat{SCA}
\end{equation}
The equation is used for each ensemble member to produce an ensemble of CF forecasts.
In the figure we see that for the years with the highest observed CF values (1994, 1999, 2006) the forecasts indicate increased likelihood of above normal CF. Conversely, for the years with less generation (2005, 2008, 2009) forecasts indicate enhanced probabilities of below normal CF.\\ 

\begin{figure}[ht!]
\includegraphics[width=\textwidth]{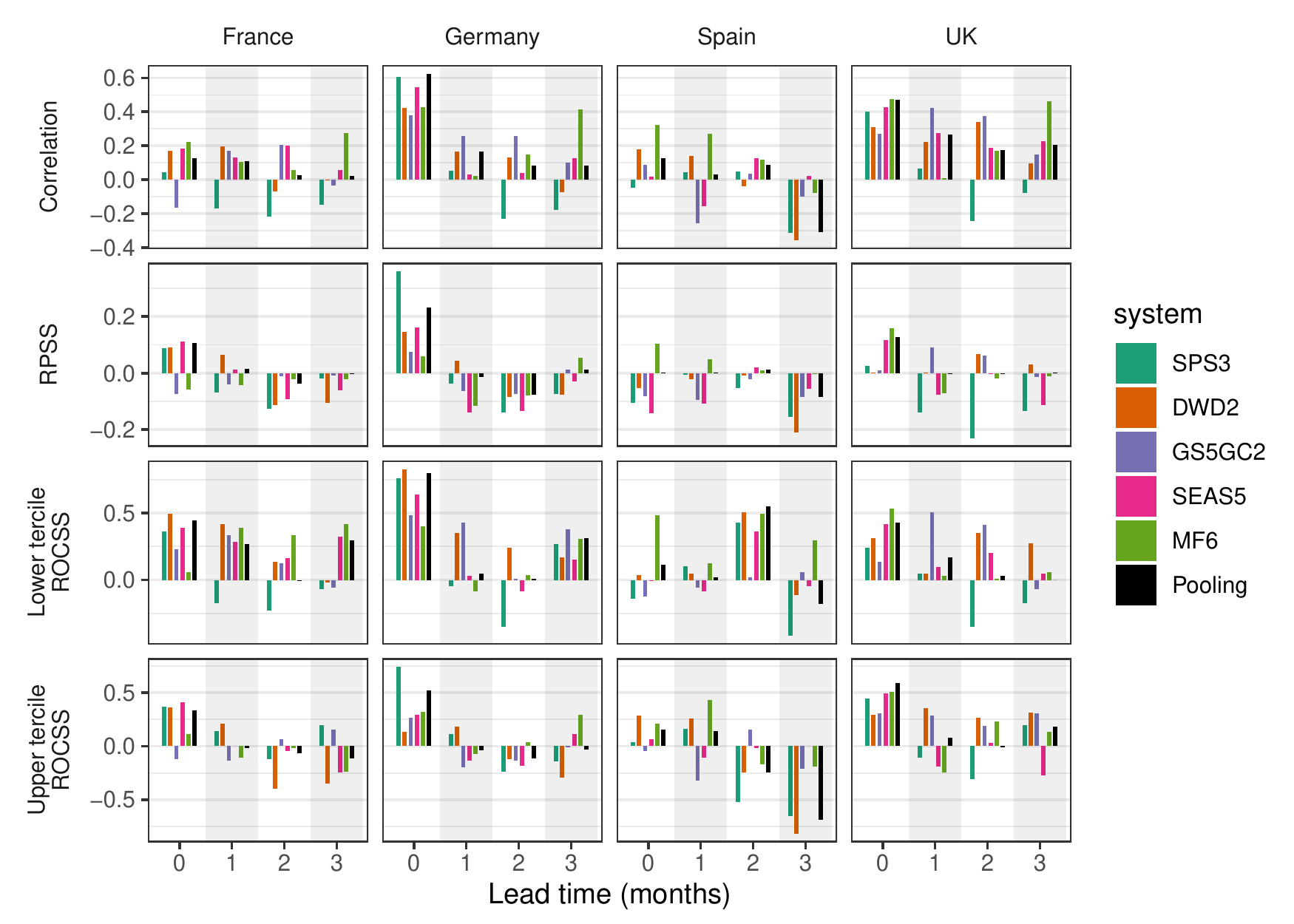}
\caption{Selected verification metrics (rows) for four European countries (columns) showing the performance of the different prediction systems at anticipating wind energy variability in winter (DJF), from zero to three months of lead time.}\label{fig:verif_series}
\end{figure}

\added{Besides this illustration, a systematic forecast quality assessment is needed to understand the performance of the predictions.}
In view of the results presented in the previous sections, we focus the analyses on wind power CF forecasts for Germany, the UK, Spain and France in winter. 
Figure~\ref{fig:verif_series} presents EMC, RPSS and ROCSS for the lower and upper terciles (see Methods for details).\\

As expected, the best results are seen for Germany and the UK, where the $r^2$ values of the regression where higher than 75\%. At lead zero, the multi-system pooling achieves an RPSS value of \replaced{0.23}{ 0.2} in Germany and \replaced{0.13 }{0.15} in the UK. These can be considered good skill levels for an RPSS value. 
Similarly, at this lead time the discrimination for the lower and upper terciles is clearly positive for all the prediction systems in Germany and the UK. 
At longer lead times there are still windows of opportunity for these two countries: at lead time three (i.e. predictions initialized on 1st September) the discrimination is positive for the lower tercile in Germany, and for the upper tercile in the UK. 
In France, all the skill scores are generally lower, with the exception of the discrimination for the lower tercile (i.e. the ability to discriminate low wind power generation winters), which is positive for most of the systems in all the lead times.
Spain presents the worst results of the four countries, with negative RPSS values for almost all systems and lead times. There is only consensus among systems in a positive value of ROCSS for the lower tercile at lead two.\\

Regarding the different SPS analyzed, there is not a single model that clearly outperforms the others for all countries and lead times, and therefore employing the multi-system is an optimal approach. Only the results for SPS3 are worth mentioning. At lead zero this system produces good results (sometimes the best results), whereas at longer leads its performance falls behind the other systems. 

\section{Discussion and conclusions}
\label{sec:discussion}
In this study we show that year-to-year changes of the large-scale atmospheric circulation over Europe impact VRE generation at country level, and we propose a way forward to anticipate generation anomalies months ahead. This is of especial importance for the future European electricity systems, that will rely on large amounts of renewable capacity and can be stressed by atmospheric conditions that are uncertain due to both natural variability and climate change effects \citep{Fabiano2021}.
The state of the general atmospheric circulation over Europe has been summarized through four Euro-Atlantic teleconnection indices. The state of those teleconnections is strongly linked with wind and solar power generation anomalies across Europe. Each teleconnection has its own areas of influence, being different for wind and solar and also changing from one season to another. This supports the idea that the most common variations of the large-scale circulation over Europe at seasonal scales, represented by the state of the teleconnection indices, produce opposite effects on wind and solar generation in many regions. Therefore, diversifying the sources of generation is a wise strategy to mitigate the risk of supply-demand imbalance at seasonal timescales. Additionally, the influence areas of the teleconnections are quite large, affecting simultaneously many neighbouring countries. This implies that at seasonal scales, surplus or lack of generation in neighbouring countries is correlated and cannot be balanced efficiently with the country interconnections. 
Similar conclusions had already been described at weekly timescales in \citet{Grams2017} and are extended here to seasonal scales. 
\deleted{We also find that current levels of solar capacity in Europe are unable to balance wind variability at seasonal scale. This puts the focus of our study on anticipating wind power generation.} \\

A multi-linear combination of the four teleconnection indices is able to represent up to 75--80\% of the winter wind generation in many northern European countries, such as UK, Ireland, Germany, Sweden or Poland. In these countries, the large-scale circulation is responsible for most of the variations in wind power generation. In summer, the situation is very different, and in most countries the teleconnections do not provide a good explanation of the interannual variations of the wind generation. In these cases, the generation might be affected by more local atmospheric conditions that are not captured by any of the four teleconnections. 
The winter teleconnections also account for a good fraction of the solar generation variability in southern Europe, whereas in summer only the generation in countries bordering the Baltic Sea is well described by the teleconnections.\\

By analyzing historical records of CF at country level we have seen that seasonal means of wind power generation are much more variable than those for solar power generation, especially in winter, which is the windiest and most variable season. Therefore anticipating winter wind power generation is of major importance in most European countries, even more if we consider that installed wind capacity almost doubles that of solar in Europe.
Luckily, the season for which forecasts of wind power are more relevant is also the season for which the teleconnection indices have better explanatory power. This might not be a mere coincidence: the largest country-wise variations of wind resource are related to large-scale physical processes that can be described by the teleconnections, whereas smaller-scale processes can have opposite effects within a single country and compensate each other, leading to smaller generation variability at country level \citep{Pickering2020}. \\

A bridging method has been applied to transform seasonal forecasts of the four winter teleconnection indices into wind capacity factor forecasts for the countries with the highest installed wind capacity, namely UK, Germany, France and Spain.
A retrospective forecast quality assessment shows that there are some windows of opportunity, especially for Germany and UK, but also France, in which forecasts of wind generation can be skillful. The quality of these predictions is mainly determined by two aspects: the goodness of fit of the multi-linear regression, and the quality of the teleconnection index forecasts \citep{Lledo2020}, which is sometimes limited. 
We have shown that not only the NAO is relevant to shape renewable generation in Europe, and in many countries the state of the EA, EAWR and SCA are key contributors too. In the countries where the multi-linear model fits well the generation, improving the seasonal forecasts of these large-scale indices would be sufficient to obtain better generation forecasts. There are still gains to be made in that direction by improving the dynamical SPS in regions where there are known biases (e.g. \citet{Exarchou2021}) and by studying the climatic processes that lead to positive/negative phases of these teleconnections. \added{Another potential way to improve the results of the final CF forecasts would be to consider the skill of the teleconnection forecasts when selecting the predictors (e.g. by giving priority to the teleconnections that can be more skillfully predicted), but implementing this remains an open challenge still.} \\

The operational provision of seasonal forecasts of those teleconnections should also be the target of future climate service developments.
From a user point of view, having forecasts of VRE generation (and also of demand and hydropower) that are derived from the large-scale circulation can be more explanatory than using separate forecasts for each atmospheric or energy variable of interest. Instead of just saying "wind power generation in Germany is likely to be above normal next winter", a user can have more confidence on a forecast formulated like this: "a positive phase of the NAO is likely in the next three months, and this is known to be accompanied with above normal winds and increased wind generation in Germany", because it describes the reasons behind in a story-line fashion that is physically self-consistent \citep{Shepherd2019}. Shaping the forecasts in this way has also the additional benefit of preserving the links between the generation of nearby regions and of complementary sources. Following with the example above, a positive NAO is also usually accompanied with positive anomalies of solar generation in Germany, and with positive anomalies of wind generation in most neighbouring countries.
A complete understanding of the supply-demand stress of the European electricity system can be gained by just employing the forecasts of the four teleconnection indices. This is a considerable reduction of degrees of freedom compared to looking at individual forecasts for each country and relevant variable and combining its uncertainties. This is indeed the realization that most of atmosphere-related stresses that the supply-demand balance might face at seasonal scale are ultimately related to the large-scale circulation, which can be faithfully monitored and anticipated employing the teleconnection indices, especially in winter.

\section*{Acknowledgments}
The research leading to these results has received funding from the European Union’s Horizon 2020 research and innovation programme under grant agreement nº 776787 (S2S4E) and nº 690462 (ERA-net ERA4CS INDECIS), and the MICINN grant BES-2017-082216 ("Ayudas para contratos predoctorales").
The authors acknowledge the Copernicus Climate Change Service (C3S) for providing seasonal predictions from several European meteorological centers and the ECMWF for producing the ERA5 reanalysis. 
We thank Stefan Pfenninger and Iain Staffell for providing the NINJA dataset, and Hannah Bloomfield, David Bryshaw and Andrew Charlton-Perez for producing the UREAD-ERA5 dataset. We acknowledge the Knowledge Management Unit, Directorate C Energy, Transport and Climate, Joint Research Centre, European Commission for the dissemination of EMHIRES.
All the analyses have been done with the R language employing the packages s2dverification\added{, CSTools,} and SpecsVerification. We want to thank Pierre-Antoine Bretonnière and Margardia Samsó for providing support with the download and formatting of the datasets.

\section*{Data availability}
The data supporting the findings of this study are openly available from different sources. 
The EMHIRES CF datasets can be obtained at  \url{https://doi.org/10.5281/zenodo.4803352}.
The NINJA CF dataset can be obtained at \url{https://www.renewables.ninja/downloads}.
The UREAD-ERA5 CF datasets are available at \url{https://doi.org/10.17864/1947.273}.
The ERA5 geopotential height data can be obtained from the Climate Data Store of the C3S \url{https://doi.org/10.24381/cds.bd0915c6is}.
The seasonal predictions of geopotential height can be accessed at \url{https://cds.climate.copernicus.eu}.

\section*{Author contributions}
LL designed the experimental setting, produced the code, analyses and figures, and prepared the manuscript. JR designed and coded the feature selection algorithm and contributed in the development of the multi-linear regression method. All authors reviewed the manuscript and contributed to the interpretation of the results.

\bibliography{main.bib}

\end{document}